
%
%
\message{Extended math symbols.}

\ifx\oldzeta\undefined    
  \let\oldzeta=\zeta    
  \def\zzeta{{\raise 2pt\hbox{$\oldzeta$}}} 
  \let\zeta=\zzeta    
\fi

\ifx\oldchi\undefined    
  \let\oldchi=\chi    
  \def\cchi{{\raise 2pt\hbox{$\oldchi$}}} 
  \let\chi=\cchi    
\fi



\def\frac#1#2{{#1 \over #2}}

\def\half{\ifinner {\scriptstyle {1 \over 2}}
   \else {1 \over 2} \fi}


\def\simge{\rlap{\raise 2pt \hbox{$>$}}{\lower 2pt \hbox{$\sim$}}}
\def\simle{\rlap{\raise 2pt \hbox{$<$}}{\lower 2pt \hbox{$\sim$}}}



\def\slashchar#1{\setbox0=\hbox{$#1$}  
   \dimen0=\wd0     
   \setbox1=\hbox{/} \dimen1=\wd1  
   \ifdim\dimen0>\dimen1   
      \rlap{\hbox to \dimen0{\hfil/\hfil}} 
      #1     
   \else     
      \rlap{\hbox to \dimen1{\hfil$#1$\hfil}} 
      /      
   \fi}      %







  \def\za{\alpha}          
             
            \def\zD{\Delta}

  \def\zg{\gamma}

  \def\zm{\mu}
  \def\zn{\nu}             
            
  \def\zp{\pi}             
              \def\zQ{\Psi}
  \def\zr{\rho}            
  \def\zs{\sigma}          
  \def\zt{\tau}
  
            \def\zW{\Omega}
  
  \def\zy{\eta}


\def\fpi{f_\pi}
\def\mpi{m_\pi}
\def\msi{m_\zs}

\def\vpi{\vec\pi}

\def\rnm{\varrho_{nm}\,}

\def\wlog#1{} 
\catcode`\@=11

\outer\def\rthnum#1{\topright{RUB-TPII-#1}}

\def\wlog{\immediate\write\m@ne} 
\catcode`\@=12 


\def\AP{\journal{Ann.\ \Phys}}

\def\NC{\journal{Nuov.\ Cim.\ }}

\def\PRep{\journal{\Phys Rep.\ }}

\def\RMP{\journal{\Rev Mod.\ \Phys}}

\def\PTP{\journal{Prog.\ Theor.\ \Phys}}
\def\PTPS{\journal{Prog.\ Theor.\ \Phys \ Supl.}}

\def\ZPA{\journalp{Z.\ \Phys}A}

\def\APPB{\journalp{Acta \ \Phys \ Polonica}B}
%
%
\input phys
\english
\titlepage
\chapters 
\equfull
\figpage
\tabpage
\refpage
\overfullrule=0pt
\FIG\Figr4{The constituent quark mass $M$ (upper part) and the sigma and
pion masses $\msi$ and $\mpi$
(lower part)
as functions of temperature at different medium densities.}
\FIG\Figr7{T-$\rho$ baryonic phase diagram. The solid curve is the
critical line for the chiral symmetry restoration. The dashed one shows
the critical line for the baryonic phase.}
\FIG\Figr8{The nucleon mass as a function of the temperature at
different medium densities.}
\FIG\Figr9{The proton charge r.m.s.radius as a function of the
temperature at different medium densities.}
\FIG\Figr10{The neutron charge r.m.s.radius as a function of the
temperature at different medium densities.}
\FIG\Figr11{The proton magnetic moment as a function of the
temperature at different medium densities.}
\FIG\Figr12{The neutron electric form factor as a function of the
temperature at different medium densities. The curves are given at zero
and two times $\rnm$ and the temperatures $T=0, 100$ and $180$ MeV.}
\FIG\Figr13{The proton electric form factor as a function of the
temperature at different medium densities. The curves are given at zero
and two times $\rnm$ and the temperatures $T=0, 100$ and $180$ MeV.}
\FIG\Figr14{The proton magnetic form factor as a function of the
temperature at different medium densities. The curves are given at zero
and two times $\rnm$ and the temperatures $T=0, 100$ and $180$ MeV.}
\FIG\Figr15{The axial form factor as a function of the
temperature at different medium densities. The curves are given at zero
and two times $\rnm$ and the temperatures $T=0, 100$ and $180$ MeV.}
\FIG\Figr16{The pion nucleon form factor as a function of the
temperature at different medium densities.}

\RF\Ripka89{G.Ripka, M.Jaminon and P.Stassart, Lecture delivered at the
Institute d'Etudes Scientifiques de Carg\`ese at the School on ``Hadrons
and Hadronic Matter'', August 8-18, 1989}
\RF\Jon73{H.F. Jones and M.D. Scadron, \AP81(1973)1*}
\RF\Eri88{T.E.O.Ericson and W.Weise, ``Pion and Nuclei'', Oxford 1988}
\RF\Fukugita88{for reviews, see M.Fukugita, {\NPB
(Proc.Suppl.)4(1988)105*}; {\NPB (Proc.Suppl.)9(1989)291*}; A.Ukawa,
{\NPB (Proc.Suppl.)10A(1989)66*}; \NPA498(1989)227c*}
\RF\Gas87{J.Gasser and H.Leutwyler, {\PLB184(1987)83*};
{\PLB188(1987)477*}; {\PLB184(1987)83*}; P.Gerber and H.Leutwyler,
\NPB321(1989)387*}
\RF\Nam61{Y.Nambu and G.Jona-Lasinio, \PR122(1961)354*}
\RF\Gell60{M.Gell-Mann and M.L\'evy, \NC16(1960)705*}
\RF\Hat85{T.Hatsuda and T.Kunihiro, {\PRL55(1985)158*}; \PTP74(1987)765*}
\RF\Hat87{T.Hatsuda and T.Kunihiro,  \PTPS91(1987)284*}
\RF\Bar89{A.Barducci, R.Casalbuoni, S.de Curtis, R.Gatto and G.Pettini,
\PLB231(1989)463*}
\RF\Cho74{A.Chodos, R.L.Jaffe, C.B.Thorn and V.Weisskopf,
{\PRD9(1974)3471*}; A.Chodos, R.L.Jaffe, K.Johnson and C.B.Thorn,
\PRD9(1974)3471*}
\RF\Gra75{T.DeGrand, R.L.Jaffe, K.Johnson and J.Kiskis,
\PRD12(1975)2060*}
\RF\Ber87{V.Bernard, Ulf-G.Meissner and I.Zahed, \PRD36(1987)819*}
\RF\Ber88{V.Bernard and Ulf-G.Meissner, \NPA489(1989)647*}
\RF\Jaminon89{M.Jaminon, G.Ripka and P.Stassart, \NPA504(1989)733*}
\RF\Sou90{K.Soutome, T.Maruyama and K.Saito, \NPA507(1990)731*}
\RF\Asa89{M.Asakawa and K.Yazaki, \NPA504(1989)668*}
\RF\Ellis89{J. Ellis, U. Heinz and H. Kowalski, \PLB233(1989)223*}
\RF\Ber89a{V.Bernard and Ulf-G.Meissner, {\PLB227(1989)465*};
\AP206(1991)50*}
\RF\Dey85{J.Dey, M.Dey and P.Ghose, \PLB165(1985)181*}
\RF\Mei89{Ulf-G.Meissner, {\PRL62(1989)1013*}; {\PLB220(1989)1*};
\NPA503(1989)801*}
\RF\Ber87a{V.Bernard  Ulf-G.Meissner and I.Zahed, \PRL59(1987)966*}
\RF\Cle86{J.Cleyman, R.V.Gavai and E.Suhonen, \PRep130(1986)217*}
\RF\Cle89{J.Cleyman, A.Koci\'c and M.D.Scadron, \PRD39(1989)323*}
\RF\Bai85{D.Bailin, J.Cleymans and M.D.Scadron, \PRD31(1985)164*}
\RF\Ait85{I.J.Aitchison and C.M.Fraser, \PRD31(1985)2605*}
\RF\Tma89{T.Meissner, E.Ruiz Arriola and K.Goeke, \ZPA336(1990)91*}
\RF\TMeissner90{T.Meissner, F.Gr\"ummer and K.Goeke, \AP202(1990)297*}
\RF\Christov90{Chr.V.Christov, E.Ruiz Arriola and K.Goeke,
{\PLB225(1989)22*}; {\NPA510(1990)1990*}; \PLB243(1990)333*}
\RF\Christov90t{Chr.V.Christov, E.Ruiz Arriola and K.Goeke,
\PLB243(1990)191*}
\RF\Christov91{Chr.V.Christov, E.Ruiz Arriola and K.Goeke,
Lecture presented at the XXX Cracow School of Theoretical Physics, June
2-12, 1990, Zakopane, Poland, \APPB22(1991)187*}
\RF\Schaldach91{J. Schaldach, E. Thommes, Chr.V.Christov and K.Goeke, DFG
Spring Meeting Salzburg 1992, February 24-28,1992, p.115}
\RF\Shi79{M.A.Shifman, A.J.Vainshtein and V.J.Zakharov,
\NPB147(1979)385,448*}
\RF\Eguchi76{T. Eguchi, \PRD14(1976)2755*}
\RF\Ber74{C. Bernard, \PRD9(1974)3312*}
\RF\Dol74{L. Dolan and R. Jackiw, \PRD9(1974)3320*}
\RF\Lutz92{M.Lutz, S.Klimt and W.Weise, \NPA542(1992)531*}
\RF\Brown90{F. R. Brown, F. P. Butler, H. Chen, N. H. Christ, Zh. Dong,
W. Schaffer, L. I. Unger and A. Vaccarino, \PLR65(1990)2491*}
\RF\Fukugita90{M. Fukugita, H. Mino, M. Okawa and A. Ukawa,
\PRL65(1990)816*}
\RF\Christ91{N. H. Christ, \NPA527(1991)539c*}
\RF\Wak80{M.Wakamatsu and A.Hayashi, \PTP63(1980)1688*}
\RF\Kaw81{S.Kawati and H.Miyata, \PRD23(1981)3010*}
\RF\Ain87{T.L.Ainsworth, E.Baron, G.E.Brown,J.
Cooperstein and M.Prakash, \NPA464(1987)740*}
\RF\Birse86{M.C.Birse, {\PRD33(1986)1934*};  Th.Cohen and W.Broniowski,
\PRD34(1986)3472*}
\RF\Fio88{M.Fiolhais, K.Goeke, F.Gr\"ummer and J.N.Urbano,
{\NPA481(1988)727*}; P.Alberto, E.Ruiz Arriola, M.Fiolhais, F.Gr\"ummer,
J.Urbano and K.Goeke, {\PLB208(1988)75*}; \ZPA336(1990)449*}
\RF\Sch51{J.Schwinger,\PR82(1951)664*}
\RF\Pau49{W.Pauli, F.Villars, \RMP21(1949)434*}
\RF\Reinhardt88{H. Reinhardt and R. W\"unsch, {\PLB215(1988)577*};
D. I. Diakonov, V. Yu. Petrov and P. V. Pobilitsa, {\NPB306(1988)809*};
T.Meissner, F.Gr\"ummer and K.Goeke, {\AP202(1990)297*};
K.Goeke, A.G\'orski, F.Gr\"ummer, Th.Mei\ss ner, H. Reinhardt
and R. W\"unsch, {\PLB256(1991)321*};
M.Wakamatsu and H. Yoshiki, {\NPA524(1991)561*};
A.G\'orski, F.Gr\"ummer and K.Goeke, \PLB278(1992)24*}

Nucl.Phys.A (1993)
\rthnum{06/91}

\title{Nucleon Properties and Restoration of Chiral Symmetry at Finite\nl
Density and Temperature in an Effective Theory}

\author{Chr.V.Christov{\rm\footnote{$^{\ddagger)}$}{Permanent
address: Institute for Nuclear Research and Nuclear Energy, Sofia 1784,
Bulgaria}}, E. Ruiz Arriola and K.Goeke}

\address{Institut f\"ur Theoretische Physik II, Ruhr-Universit\"at
Bochum, D-4630 Bochum}

\abstract{Modifications of baryon properties due to the restoration of the
chiral symmetry in an external hot and dense baryon medium are investigated
in an effective chiral quark-meson theory. The nucleon arises as a
soliton of
the Gell-Mann - L\'evi $\zs$-model, the parameters of which are chosen to be
the medium-modified meson values evaluated within the Nambu -
Jona-Lasinio model. The nucleon properties are obtained by
means of variational projection techniques. The nucleon form factors
as well as the nucleon delta transition form factors are evaluated for
various densities and temperatures of the medium. Similar to the chiral
phase transition line the critical curve in the $T-\zr$ plane for
delocalization of the
nucleon is non-monotonic and this feature is reflected in all nucleon
properties. At medium densities of about $(2-3) \rnm$ the baryonic
phase exists only at intermediate temperatures. For finite temperature
and densities the nucleon form factors get strongly reduced at finite
transfer momenta.}

\endpage

\chap{Introduction}
The QCD as a theory of the strong interaction is assumed to
incorporate the spontaneous chiral symmetry breaking.
The latter is believed to be a dominant mechanism in the low-energy sector
of QCD and in particular, for the structure of the low-lying baryons.
As it is suggested by the lattice QCD
calculations\quref{\Fukugita88}),
at some finite temperature or/and at some finite density one generally
should expect a restoration of the chiral symmetry.
Hence, it is a common idea to consider the (partial) restoration of the
chiral symmetry in a hot and dense medium as a relevant mechanism for a
scale change which modifies the structure of the baryons immersed in it.

Since on one hand chiral models like Nambu --
Jona-Lasinio\quref{\Nam61}) and Gell-Mann -- L\'evy
$\zs$-model\quref{\Gell60}) account for the chiral symmetry breaking
and on the other they allow for the (partial) restoration of the symmetry
at finite density and/or temperature those models seem to provide a
suitable working scheme to study modifications of the meson and baryon
properties in medium. It should be noted also that the critical value
(of about 200 MeV) for the temperature is much lower than the typical
cutoff used to regularize the NJL model. Indeed, the models show a
significant success in description
of both the static and dynamic properties of the nucleon arising as a
non-topological soliton\quref{\Birse86\use{\Fio88}-\Reinhardt88}). Quite
encouragingly, assuming that the nuclear medium can approximately be replaced
by a uniform quark medium, the investigations
\quref{\Hat85\use{\Ber87\Cle89\Asa89\Christov90t}-\Lutz92}) based on these
models give a restoration at both finite temperature and density in a
quantitative agreement with the Monte-Carlo lattice calculations as well
as with the chiral perturbation theory\quref{\Gas87}).
Medium\quref{\Ber87,\Mei89\use{\Jaminon89}-\Christov90})
and temperature\quref{\Hat87,\Ber89a}) effects in the properties of the
mesons and of the nucleon have been successfully studied as well. Despite of
that the meson and nucleon properties seem to be affected by the density and
by the temperature in a similar way, the physical
mechanism, however, driving the chiral symmetry restoration in each
case, is different. At finite density the attractive interaction of the
medium with the Dirac sea polarize it in a way that the quark condensate
$<\overline qq>$ gets reduced whereas at finite temperature the thermal
fluctuations simply disorder the system. Therefore one might expect
non-trivial effects to meson and baryon properties if both temperature and
density are assumed finite. Indeed, the meson
properties\quref{\Lutz92,\Christov91}) show a non-trivial temperature
dependence at finite density.  It is illustrated in \qufig{\Figr4}
a) and b) where the constituent quark mass and the meson masses
calculated\quref{\Christov91}) within the NJL
model are shown for four different baryon densities
as a function of temperature.
We also studied\quref{\Christov90t}) the modification of
some static nucleon properties, namely the mass and the charge radius, due
to the chiral symmetry restoration in hot and dense baryon medium. The
nucleon appears as a soliton in the Gell-Mann -- L\'evy $\zs$-model
which is uniquely defined by medium-modified meson values coming from
the NJL model. We found a non-monotonic critical curve for delocalization of
the nucleon in the medium. The nucleon mass and the charge radius show also
a non-monotonic temperature dependence at finite density.

In the present work we extend the study of ref.\quref{\Christov90t}) to
nucleon form factors. Thus, the main task is to study in a systematic
way the modification of the
nucleon properties due to the restoration of the chiral symmetry in
dense and hot medium. This is done by solving the medium-modified
Gell-Mann -- L\'evy $\zs$-model using angular and isospin projection
combined with variational techniques. Various form factor and observables
of the nucleon and the delta-isobar are evaluated.

\chap{The projected chiral soliton model}

In our approach we assume a physical picture in which the nucleon, being
off-shell by the presence of the medium, is simulated by an on-shell
nucleon facing meson fields modified by the medium. In fact it means
that the influence of the medium is expressed in terms of modified
values of the pion decay constant, and the pion and sigma masses. To that
end we use the NJL model in an approximation which consists of
treating the baryon medium as an uniform quark matter neglecting the
nucleon sub-structure in it. The modified values of $\fpi$, $\mpi$ and
$\msi$ are then used to define a Gell-Mann -- L\'evy $\zs$-model for the
nucleon (in medium) which is later solved by standard self-consistent
methods.

We determine the properties of the nucleon employing the
projected chiral soliton model\quref{\Fio88}). Here we briefly
sketch its main points. It is given by the Gell-Mann--L\'evy
lagrangean\quref{\Gell60}) with valence quarks solved by a variational
procedure with a spin and isospin projection.

The Gell-Mann - L\`evy lagrangean reads:
$$
{\cal L}=\overline\Psi [i\zg^\zm\partial_\zm-
g(\zs+i\zg_5\vec\zt.\vpi)]\zQ+\frac 12\partial^\mu\sigma\partial_\mu\sigma
+\frac 12 \partial^\mu\vpi\partial_\zm\vpi-U(\sigma,\vpi).
\EQN\Eq27
$$
with the meson self-interaction potential
$$
U(\sigma,\vpi)=\frac {\zy^2}2(\zs^2+\vpi^2-\zn^2)^2+\fpi\mpi^2\zs.
\EQN\Eq28
$$
Besides the coupling constant $g$ the others can be expressed by means
of $\mpi$, $\msi$ and $\fpi$:
$$
\zy^2=(\msi^2-\mpi^2)/2\fpi^2, \qquad \hbox{ and }\qquad
\zn^2=(\fpi^2-\mpi^2)/\zy^2. \EQN\Eq29
$$
In the present approach at finite density and temperature
the lagrangean is simply modified by inserting their medium values.
Actually the NJL model can be related to the
Gell-Mann--L\'evy sigma model via a gradient expansion and hence to
combine these two models, as it is done here, seems to be reasonable.

In order to get a solitonic solution of lagrangean \queq{\Eq27} we
employ a variational procedure based on mean-field states with a
generalized hedgehog structure and spin and isospin projection
(described in detail in ref.\quref{\Fio88}). The trial functions for
the nucleon is assumed to be product of coherent Fock states
representing the sigma and pion fields and a quark state
taken to be an antisymmetrized product in
colour space of three valence quarks in the same 1s-orbit.
The states of a good angular momentum and isospin are
obtained from the classical solitonic solution by means of
projection operators of the type:
$$
P_{MK(M_TK_T)}^{J(T)}\sim \int d{\bf \zW}\,D_{MK(M_TK_T)}^{J(T)*}\hat
R({\bf \zW}), \EQN\Eq31
$$
where $JM$ ($TM_T$) are spin (isospin) numbers and $\za$ stands for the
additional quantum numbers. $R({\bf \zW})$ is the rotation operator in spin (or
isospin) space
and ${\bf \zW}$ denotes the set of Euler angles in this space.

The obtained projected nucleonic solution with good spin and isospin
numbers is used to evaluate the static nucleon properties as well as
the nucleon and the nucleon-delta transition form factors. The form
factors are evaluated in the Breit frame neglecting the recoil effects
and assuming that the projected state of the nucleonic soliton is a
good zero momentum state. In the case of the nucleon-delta transition
form factors we work with the Rarita-Schwinger formalism in the limit of
nucleon-delta mass degeneracy. One can find the details in
refs.\quref{\Fio88}). For completeness we present the final
expression for the form factors in terms of the projected state:
$$
G_E(q^2)=\int d^3r\,j_0(qr)<JTMM_T|j^0_{em}({\bf r})|JTMM_T>, \EQN\Eq32
$$
$$
\frac {G_M(q^2)}{2M_N}=\frac 32 \int d^3r\,\frac
{j_1(qr)}{qr}[{\bf r}\times<JTMM_T|{\bf j}_{em}({\bf r})|JTMM_T>,
\EQN\Eq33
$$
$$
\eqalign{
G_A(q^2)=&\frac {2M_N}{\sqrt{q^2/4+M_N^2}} \int
d^3r\,\{j_0(qr)<JTMM_T|A^3_z({\bf r})|JTMM_T>\cr
& -\sqrt{2\pi}j_2(qr)[Y_2\otimes <JTMM_T|{\bf A}^3_z({\bf
r})|JTMM_T>]_{10}\},\cr} \EQN\Eq34
$$
$$
G_{\pi NN}(q^2)=-6M_N \int d^3r\,\frac
{j_1(qr)}{qr}z<JTMM_T|\vec j_\pi^3({\bf r})|JTMM_T>. \EQN\Eq35
$$
$$
{G^M_{N\zD} \over 2 M_N}= \sqrt{6} \int d^3r {j_1(qr)
\over qr} <\Delta ^+_ {1 \over 2}
|\hat{\mu}_0|N^+_{1 \over 2} >,\EQN\Eq36
$$
$$
G_{\pi N\Delta}= 6 M_N \frac {M_N^2}{q^2/4+M_N^2}\frac 32 \int d^3r
{j_1(qr)\over qr} z <\Delta ^+_ {1 \over 2}
|\widehat{J}^0_\pi|N^+_{1 \over 2} >. \EQN\Eq37
$$
Eqs.\queq{\Eq32} and \queq{\Eq33} are the electromagnetic Sachs form
factors as the ${\bf j}^\zm_{em}$ is the $\zm$-component of the
electromagnetic current operator. The axial form factor
is given by eq.\queq{\Eq34} where the $A^\zm_i$ is the $i$-space and
$\zm$-isovector component of the axial current operator. Eq.\queq{\Eq35}
concerns the pion nucleon form factor evaluated by the pion source
current $\vec{\bf j}_\zp$. For the $N-\zD$ transition magnetic form
factor \queq{\Eq36} we use the standard decomposition for the matrix
element of the electromagnetic current given by Jones and
Scadron\quref{\Jon73}). For the $\pi N\Delta$ form factor
$G_{\pi N\Delta}$ \queq{\Eq37} the simplest coupling for the $\pi N\Delta$
vertex\quref{\Eri88}) is assumed.

\chap{Nucleon as a soliton at finite density and temperature}

In the present approach the nucleon appears as a
self-consistent localized stationary solution (soliton) in a suitable
modified Gell-Mann - L\'evy sigma model solved in a variational
procedure with a spin and isospin projection.
The first problem we would like to look into is the limits which the
restoration of the chiral symmetry in a hot
and dense medium imposes on the existence of the nucleon as a soliton.
In this model picture the lack of solitonic
solution is taken as an indication for a delocalization of the baryonic
phase. It is rather tempting to identify this with deconfinement but a
such conclusion seems to be out of the scope of the model itself.
\qufig{\Figr7} shows the resulting critical line for existing of the
nucleon in the $T-\zr$ plane together with the chiral phase
diagram expected\quref{\Christov91}) from the meson sector. One can see
that both critical
curves have a nearly identical behaviour where the critical values for the
baryonic phase are smaller than ones for the chiral phase transition. It
means that the delocalization of the nucleon always happens before the
restoration of the chiral symmetry takes place. Apparently the change
of the chiral symmetry properties drives the delocalization of the
nucleon. It should also be noted that for densities (2-3)$\rnm$
which are relevant for the high energy heavy-ion reactions the baryonic
phase exists only at intermediate temperatures.

\sect{Nucleon static properties at finite density and temperature}

Using the projected nucleonic solution with good spin and isospin
quantum numbers we have also calculated at finite density and
temperature all nucleon properties including the nucleon and
nucleon-delta transition form factors as well. In this section we
present the results concerning the nucleon static properties. The
calculated nucleon mass as a function of the temperature at different
densities is shown in \qufig{\Figr8}. At zero density the nucleon mass
shows no change at increasing temperatures up to $1/2\,T_c$ and
decreases then monotonically at higher temperatures. Near the critical
value $T_c$ the nucleon mass is noticeably reduced. Similar behaviour is
concluded by Bernard and Meissner\quref{\Ber89a}) within the Skyrme model
with vector mesons. It is interesting also to mention that Ellis et al.
\quref{\Ellis89}) predict that the reduction of the nucleon mass will lead
to an enhancement of the antibaryon production in heavy-ion collisions.
At finite densities, however, the temperature dependence of the nucleon
mass is quite different. For temperature
up to about $3/4 \,T_c$ the mass increases which means that the nucleon
becomes more stable. At density above the critical value $\zr_c$ the
nucleon exists only at intermediate temperatures which agrees
with the critical line shown in \qufig{\Figr7}. The delta mass as well
as the nucleon-delta mass splitting have exactly the same behaviour as
those of the nucleon mass shown in \qufig{\Figr8}.

\qufig{\Figr9} and \qufig{\Figr10} illustrate the density and
temperature dependence of the proton and neutron charge radii. Apart
from the values at the origin they have similar behaviour. At the
critical temperature value all curves show a clear trend to grow to
infinity  which indicates the delocalization of the nucleon. At
vanishing medium density both quantities increase monotonically. At
finite density values it changes as at intermediate
temperatures the radii get reduced. The latter is a clear signal for a
stabilization of the nucleon. This is very
well seen at high densities where the nucleon has a finite
radius only at intermediate temperatures. We found similar behaviour in
the magnetic and axial r.m.s.radii at finite density and temperature.

It should be emphasized that the
vanishing of the nucleon mass and the unlimited increase of the radius
happen always along the critical line for the nucleonic soliton
solution. It means that at least in present model picture the lack of
the solitonic solution should be identified with a delocalization of the
nucleon.

The proton magnetic moment calculated at finite density and temperature
is depicted in \qufig{\Figr11}. The same picture is valid for the
neutron magnetic moment as well as for $N-\zD$ transition magnetic
moment. As can be seen the magnetic moment shows a behaviour similar
to those of the charge radius but the medium effect is much less
pronounced.

In contrast to the other nucleon properties the axial vector coupling
constant and the pion nucleon coupling constant  as well as the pion
nucleon delta coupling constant show a very slight medium modification.
Apart from the critical density and temperature values they
stay almost constant. The $g_A$ and $g_{\zp N\zD}$, however, are a bit
more affected: the first quantity shows a slight decrease whereas
the second one a slight increase. For instance, at temperature
close to $T_c$ both show a common change of about 10\% relative to the
free value. At the critical temperature and density values all coupling
constants vanish rapidly. Dey et al.\quref{\Dey85}) concluded a bit
larger temperature effect in $g_{\zp NN}$. In their approach the $g_{\zp
NN}$ is bound to decrease with the temperature.
Their considerations, however, are based on a particular assumptions
about the nucleon current and are limited to relatively low temperatures
because of the approximations used.

\sect{Nucleon and nucleon-delta transition form factors at finite
density and temperature}

Using the projected nucleonic solution with good spin and isospin
numbers in eqs.\queq{\Eq32-\Eq28} we evaluate the nucleon and the
nucleon-delta transition form factors. The results for the proton and
neutron electric form factors are plotted in \qufig{\Figr12} and
\qufig{\Figr13} for a zero and two times nuclear matter density and
three different temperatures. As can be seen at zero density and
relatively low temperatures ($<100$ MeV) the effects are negligible.
At finite density and temperature the form factor gets reduced at
finite transfer momenta in comparison with the free one. One realizes,
however, that similar to other nucleon properties at intermediate
temperatures the density effects are partially suppressed by the
temperature and the form factor are less reduced than in the case of
vanishing temperature and finite density. The slope of the origin also
changes which is reflected in the corresponding r.m.s.radii. Close to
the critical values the temperature effects are dominant and strong
enough to make the form factors practically vanishing at large transfer
momenta.

One can find the same effects in the other form factors. In
\qufig{\Figr14} the proton magnetic form factor is depicted. As in the
previous case there is no change at zero density and low temperatures
up to 100 MeV. The crossing known from the finite density
study\quref{\Christov90}) is still presented. The magnetic $N-\zD$
transition form factor has an identical behaviour. The last two form
factors presented (see \qufig{\Figr15} and \qufig{\Figr16}) are the
axial and the pion nucleon form factors. The pion nucleon delta form
factor is similar to the pion nucleon one. Both seem to be more
affected by the medium at finite momentum transfers than the other form
factors.

\chap{Summary}

We have investigated the modification of the nucleon properties due to the
restoration of the chiral symmetry in a hot and
dense baryon medium within an effective chiral quark-meson theory.
The nucleon appears as a soliton in the Gell-Mann -- L\'evy $\zs$-model
which is defined in terms of the modified values
of the pion decay constant and the pion and sigma masses in the medium.
These values are obtained from the NJL model.
The critical curve in $T-\zr$ plane for delocalization of the nucleon
is non-monotonic. The delocalization occurs before
the chiral phase transition. At medium densities of about (2-3)$\rnm$
the baryonic phase exists only at intermediate temperatures. The
nucleon properties evaluated by means of projection techniques also
show non-monotonic dependence on temperature at finite density. At
critical density and temperature the nucleon mass vanishes whereas the
radii grow to infinity. At finite density and intermediate temperatures
both quantities indicate a stabilization of the nucleon. At finite
density and temperature values all form factors get reduced at finite
transfer momenta. The reduction, however, is smaller than in the case
of vanishing temperature.

{\it This work is partially supported by the Bundesministerium f\"ur
Forschung und Technologie, the KFA J\"ulich (COSY-Project) and the Deutsche
Forschunggemeinschaft}

\refout
\figout

\end